\documentclass[hidelinks,times,twocolumn,final,longtitle]{elsarticle}

\usepackage{jasr}
\usepackage{framed,multirow}

\usepackage{amssymb}
\usepackage{latexsym}

\usepackage{natbib}

\usepackage[switch]{lineno}

\usepackage{url}
\usepackage{xcolor}
\definecolor{newcolor}{rgb}{.8,.349,.1}

\usepackage[citebordercolor=white]{hyperref}
\usepackage{amsmath}

\journal{Advances in Space Research}

\begin{document}

\verso{Daniel S. Roll \textit{et al}}

\begin{frontmatter}

\title{CosmosDSR - a methodology for automated detection and tracking of orbital debris using the Unscented Kalman Filter\tnoteref{tnote1}}%

\author[1]{Daniel S. \snm{Roll}\fnref{fn1}}
\author[2]{Zeyneb \snm{Kurt}\corref{cor1}}
\author[1]{Wai Lok \snm{Woo}\corref{cor1}}
\fntext[fn1]{First author: Email: dan.roll@northumbria.ac.uk}
\cortext[cor1]{Corresponding authors: 
  Email: wailok.woo@northumbria.ac.uk, z.kurt@sheffield.ac.uk}

\affiliation[1]{organization={Northumbria University},
                addressline={College Street},
                city={Newcastle upon Tyne},
                postcode={NE1 8ST},
                country={England}}
\affiliation[2]{organization={University of Sheffield},
                addressline={Western Bank},
                city={Sheffield},
                postcode={S10 2TN},
                country={England}}

\received{1 May 2013}
\finalform{10 May 2013}
\accepted{13 May 2013}
\availableonline{15 May 2013}
\communicated{S. Sarkar}

\begin{abstract}
The Kessler syndrome refers to the escalating space debris from frequent space activities, threatening future space exploration. Addressing this issue is vital. Several AI models, including Convolutional Neural Networks, Kernel Principal Component Analysis, and Model-Agnostic Meta-Learning have been assessed with various data types. Earlier studies highlighted the combination of the YOLO object detector and a linear Kalman filter (LKF) for object detection and tracking. Advancing this, the current paper introduces a novel methodology for the \textit{Comprehensive Orbital Surveillance and Monitoring Of Space by Detecting Satellite Residuals} (CosmosDSR) by combining YOLOv3 with an Unscented Kalman Filter (UKF) for tracking satellites in sequential images. Using the Spacecraft Recognition Leveraging Knowledge of Space Environment (SPARK) dataset for training and testing, the YOLOv3 precisely detected and classified all satellite categories (Mean Average Precision=97.18\%, F1=0.95) with few errors (TP=4163, FP=209, FN=237). Both CosmosDSR and an implemented LKF used for comparison tracked satellites accurately for a mean squared error (MSE) and root mean squared error (RME) of MSE=2.83/RMSE=1.66 for UKF and MSE=2.84/RMSE=1.66 for LKF. The current study is limited to images generated in a space simulation environment, but the CosmosDSR methodology shows great potential in detecting and tracking satellites, paving the way for solutions to the Kessler syndrome.

\end{abstract}

\begin{keyword}
\KWD Unscented Kalman Filter\sep YOLO\sep Orbital Debris
\end{keyword}

\end{frontmatter}


\section{Introduction}
\subsection{Background}
As machine learning (ML) applications increase in prevalence, one area that has gained much traction over the last few decades is computer vision (CV), a field concerned with developing methods for a computer to interpret and understand visual information from images and videos. One common CV technique that involves identifying objects within such media is known as object detection (OD), combining the technology of image processing with deep learning (DL). The benefits of OD technology are numerous, including increased efficiency and performance by automating tasks that require human participation, reducing necessity of manual labour as well as being able to do so in a time-sensitive manner that is scalable to large datasets \citep{liu_2019_deep}. An increasingly significant domain that has potential to profit greatly from the application of OD is that of space – specifically in terms of orbital debris and satellite detection.

Decades of humanity’s endeavour to investigate, explore and innovate concerning the stellar environment have led to the accumulation of unwanted items in orbit, ranging in size and origin from millimetre-wide flecks of paint eroded from spacecraft hulls to intact dysfunctional satellites \citep{NASA_2019_ares}. More than 23,000 tracked ‘resident-space objects’ (RSO’s) currently circle the Earth. Such objects are known colloquially as ‘space junk’, comprising mostly of discarded elements of humanity’s previous space ventures, with only a small percentage amounting to functional satellites or spacecraft \citep{kaineg_2020_the}. Such items pose serious risks for the future of both space investigation and travel. Research estimating the speed of collisions involving pieces of orbital debris posited possible speeds of up to 50,000 mph \citep{liou_2006_risks}, not only fatal to life, but in the event of a collision with another object, catalysing the production of further debris fragments from the resulting impact. Thus, if thorough analysis and cataloguing of orbital debris is not rigorously maintained, such risks will increase exponentially in line with the degradation of the celestial environment. Known as The Kessler Syndrome \citep{kessler_1978_collision}, this idea of cascading collisions has the potential to not only severely delay future space travel, but even end it entirely. The accumulation of such objects, therefore, must be addressed. One common issue, however, is the necessity for accurate measurements of information relevant to approximating the relative positions of debris fragments. Accomplishing this could incorporate the design and implementation of AI technologies and as shown in the literature, such intelligent solutions have already been theorised and tested. Currently, this includes the use of ResNet \citep{perez_detection}; customised traditional CNN models \citep{jahirabadkar_2020_space}; KPCA with 2-D wavelet transformations \citep{ma_2011_space} and MAML \citep{furfaro_space}. Such attempts are varied, and target vastly different sub-domains within orbital debris; yet work together to build a picture of the scale and complexity of the problem. Understandably, with such variation, the data employed in the studies is also varied, including light-curve estimations \citep{allworth_2021_a}, RGB images \citep{aldahoul_2022_rgbd} and radar scans \citep{mehrholz_2002_detecting}, with image data being the most common. The possibility to use such images in conjunction with CV technologies such as OD offers the ability to design and test bespoke software that can subsequently be tested and deployed using sporadic real data. One piece of previous literature that applies contemporary OD to identify space objects is research by Fitzgerald \citep{fitzgerald_2022_space} to determine objects in low-resolution, wide field-of-view (WFOV) synthetic images of the night sky representative of physical images taken by the PANDORA sensor array located at the Air Force Maui Optical and Supercomputing Site in the USA. We propose architecture, in the form of an object detector and tracker using a blend of YOLO and a Kalman Filter (KF), to be deployed to autonomously monitor space objects for cataloguing and anomaly detection purposes.

\subsection{You Only Look Once (YOLO}
The YOLO algorithm used was first introduced by \citet{redmon_2016_you} in 2016 as an improvement to its contemporaries with regards to speed and performance. Whilst previous two-step detector algorithms such as R-CNN require their nominal two steps, firstly identifying regions of interest and then classifying those regions \citep{carranzagarca_2020_on}, YOLO revolutionised this with the ability to process images in a single forward pass. This is achieved by applying a single neural network to the whole image, allowing for the simultaneous prediction of both bounding boxes and class probabilities. YOLO therefore proposed remarkable improvements in speed and efficiency, establishing itself as a method for real-time object detection and offering a solution for applications requiring instant feedback. Following the development of the initial YOLO algorithm, many other iterations have been published, such as YOLO9000 (v2) \citep{redmon_2016_yolo9000} and recently, YOLOv8 \citep{redmon_2018_yolov3}. Each version aimed to address limitations of the previous, offering improved class predictions, deeper architectures and more detection layers. The widespread deployment of YOLO across vast domains and topics is demonstrated, including algorithms trained to identify traffic lights for the development of autonomous vehicles \citep{mostafa_2022_a}; for cell counting in fluorescence microscopy for medical research \citep{aldughayfiq_2023_yolov5fpn}; for pest detection within agricultural settings \citep{zhang_2022_agripestyolo}. With regards to the stellar environment, versions of YOLO have already been successfully applied to various topics including residual space object (RSO) detection and tracking \citep{mastrofini_2023_yolo}, satellite component recognition \citep{mahendrakar_2021_realtime} and autonomous target detection from satellite imagery \citep{tahir_2022_automatic}. There are certain inherent drawbacks, however, such as difficulty with recognising small or overlapping objects or the impact of noise and occlusion on the detection \citep{terven_2023_a}.  

\subsection{The Kalman Filter}
The Kalman Filter is a recursive mathematical algorithm developed by the electrical engineer Rudolf E. Kálmán in the 1960’s as a method for simplifying linear filtering and prediction problems \citep{kalman_1960_a}. A KF, in terms of a system, builds upon previous output to compute future output by continuously updating its estimates dependent on new information; using prior state estimates and current data, the filter predicts the next state of the system. This allows a KF to evaluate the changing state of a linear, dynamic system whilst incorporating noise and incomplete measurements with the prior estimates, mostly negating their negative impact on estimate accuracy \citep{urrea_2021_kalman}. As such, the practical applications of KF are diverse, including time-series forecasting and stock price prediction within finance \citep{rankin_kalman, martinelli_predicting}, autonomous robot localisation \citep{zhafri_autonomous} and many fault-diagnosis or sensor-control functions within industry \citep{auger_2013_industrial}. The KF offers benefits such as optimum state estimation ability for such linear systems with Gaussian noise, as well as its efficiency in real-time scenarios, yet is therefore somewhat restricted as assumptions of linearity and Gaussian noise must be upheld \citep{kim_2018_introduction}. This, therefore, makes KF less suitable for non-linear systems or those with non-Gaussian noise and led to the development of the Unscented Kalman Filter (UKF). The UKF approximates non-linear functions using a deterministic sampling technique – the Unscented Transformation (UT) – as opposed to attempting to linearise them \citep{ronghuizhan_2006_neural}. This results in high accuracy of estimation for many applications and is therefore routinely employed in areas where such high accuracy of state estimations is vital, such as autonomous vehicle sensor fusion and other advanced navigation systems \citep{krauss_2022_unscented}. 

\subsection{Combined Approaches}
Unsurprisingly, attempts to blend the KF into deeper architecture, as a method to track detected objects in OD applications such as YOLO algorithms, have been attempted. Research by \citet{barreiros_2021_zebrafish} attempted to detect and track zebrafish within videos captured in an experimental setting reminiscent of natural marine conditions. In order to successfully track fish during rapid movement sequences, in which occlusion causes inconsistency of detections, a new approach was necessary as previous literature mainly studied shallower water or more stationary fish. The authors designed a convolutional network for their object recognition using YOLOv2 to delimit the region of a fish’s head and facilitate individual fish detection. Following this, a KF was used to estimate the head position and track the trajectory of each fish throughout subsequent frames. Results indicated superior performance of the YOLO, indicating its ability to detect fish even when presented with varying numbers or occlusion. The tracking with KF also performed well on frames with optimal characteristics, such as a low number of fish, however encountered problems and lowered accuracy when presented with a higher number, faster swimming motions or more occlusion. Overall, however, the proposed system worked to effectively detect and track zebrafish within the experimental setting. Furthermore, similar architecture has been applied to the field of space as observed in the previously mentioned study by Fitzgerald \citep{fitzgerald_2022_space}. The author applied a model consisting of a deep-learned object detector using YOLOv5 combined with KF for object tracking to synthetic images of RSO’s within the Geosynchronous Equatorial Orbit (GEO) belt representative of physical images taken by the WFOV sensors of the PANDORA system. Aiming to first process the images, to detect low-light objects, before tracking those objects as a method of passive monitoring, the model was then applied to a real, physical dataset as to evaluate performance in a tangible application. Moreover, the architecture was compared and contrasted to classical OD methods: Scale Invariant Feature Transformation (SIFT), Histogram of Oriented Gradients: Orientation Rank (HOGOR) and Form Factor (FF). The author elucidates that the proposed YOLO/KF architecture was not only efficient and accurate when performing object tracking and detection across all test sets of images, but also comparatively outperformed the classic detectors. Fitzgerald notes that whilst the research was a success, increasing GEO RSO detection accuracy from the classical methods, design improvements could be made, and future work could aim to also compare the results to other sensors involved in space object monitoring.

\subsection{Research Gap}
The available literature therefore advocates for the usage of OD methods, specifically object detection and tracking combinations, in the domain of RSO identification. The state of the stellar landscape is exponentially declining, and measures must be taken to ‘clean up’ the environment. The justification to apply the above methods to this problem is therefore substantiated, allowing for a novel design that combines leading technologies into an optimised solution for RSO classification and tracking. Whilst the research by Fitzgerald focused on the integration of the linear KF, based on the observed benefits from the literature, there exists a gap in which to probe the efficacy of the UKF for RSO tracking and integrate it in combination with YOLO architecture to facilitate identification and classification. Due to the complex, multifaceted and often non-linear nature of the orbital debris problem, incorporating an algorithm inherently capable of dealing with such non-linearity proposes many benefits. As such, this thesis outlines a model and methodology for RSO identification and tracking from synthetic images of satellites using a combination of a deep-learned object detector using YOLOv3 and an Unscented Kalman filter. The study aims to evaluate the performance of the UKF by directly comparing prediction accuracies and precision to those generated by a linear KF. Successful implementation of novel architecture would stand to fill numerous gaps within the research area, addressing some of the associated limitations brought forth by previous work and further developing the understanding of RSO tracking methodologies. This, in turn, would offer a solution to orbital debris tracking that could be further implemented into removal technologies as a solution to address the developing Kessler syndrome, stimulating stellar clean-up and expediating future space exploration.

\section{Data Selection}
\subsection{Data Evaluation}
Upon review, there existed not enough easily accessible or publicly available real RSO images to form a dataset (even with data augmentation), and it was noted that there is a distinct lack of readily available real image data regarding satellites available to researchers. Moving forward, collating a repository of real images is a must, even if this only results in a dataset for usage as a confirmation set, as to test architecture on real, empirical data. However, this was determined to be outside the scope of the current project and a decision was made to focus on synthetic datasets containing simulated, computer-generated images. Literature already supports the use of synthetic data within the field, not only the previous study by \citet{fitzgerald_2022_space} but also research by \citet{zhang_2022_a} which developed a diverse space-target dataset including both satellites and debris. The current study, however, focuses on the Spacecraft Recognition Leveraging Knowledge of Space Environment (SPARK) dataset, available from the University of Luxembourg and first published in research by \citet{musallam_spark}. 

\subsection{The SPARK Dataset}
The SPARK is a unique multi-modal space object image dataset that was developed to be used in Space Situational Awareness (SSA) applications. All images were generated under a realistic space simulation environment, with diverse sensing conditions producing diverse orbital scenarios, with the authors positing that preliminary experimental evaluation suggests the dataset is both valid and relevant. This would identify its utility in training OD applications. The dataset contains images of n=11 distinct classes of RSO, the names of which can be seen in Table 1. 

The data is split into two streams, Stream-1 and Stream-2. The first contains images in 1080x1080 resolution of all 11 classes, supplied with relevant bounding-box data. The second is for trajectory predictions and contains numerous folders each containing a set of consecutive images in 1440x1080 resolution of one RSO, usually in a much smaller format.  

\subsection{Data Retrieval and Processing}
A formal request for access to the dataset was made to the University of Luxembourg. For the current study, using the entire dataset was deemed to be excessive and a subset was created for use throughout. This resulted in a current data pool that comprised n=22000 images, equally distributed between the 11 classes, from Stream-1 as well as n=1500 images from n=5 sets from Stream-2. Example images from both streams are shown in Figure 1.  

Following successful partitioning of the dataset, annotation files were created for each image. This was streamlined with the use of a simple processing script written in R using R Studio v2021.09.0. The label files were added to the Stream-1 folder and, alongside the relevant sets from Stream-2, it was uploaded to Google drive for use within Google Colab. 

\begin{figure}
  \centering
  \includegraphics[scale=0.4]{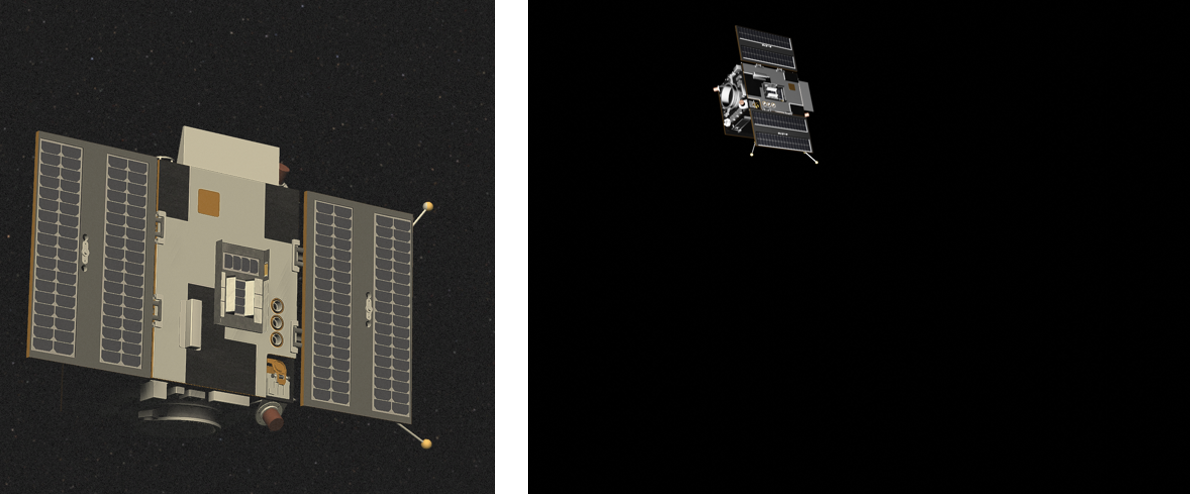}
  \caption{Examples of images from Stream-1 (LEFT) and Stream-2 (RIGHT) of the SPARK}
  \label{fig:pendulum}
\end{figure}

\begin{table}
\centering
\caption{RSO classes within the SPARK}
\begin{tabular}{|l|l|}
\hline
Class ID & Name  \\
\hline
0 & lisa\_pathfinder \\
\hline
1 & proba\_data3\_csc \\
\hline
2 & smart\_1 \\
\hline
3 & xmm\_newton \\
\hline
4 & soho \\
\hline
5 & earth\_observation\_sat\_1 \\
\hline
6 & debris \\
\hline
7 & proba\_2 \\
\hline
8 & proba\_3\_ocs \\
\hline
9 & cheops \\
\hline
10 & double\_star \\
\hline
\end{tabular}
\end{table}

\section{Algorithm Selection}
\subsection{YOLO}
As discussed above, the current study elected to utilise a YOLO algorithm for OD purposes, in part due to the wealth of available literature demonstrating its effectiveness in similar scenarios, but also due to the possibility to easily integrate a KF for object tracking. Various versions of YOLO were analysed for their suitability for use within the current study. YOLOv3 was developed and published in 2018 by \citet{redmon_2018_yolov3} as a successor to the widely popular YOLO9000. It offered a larger overall architecture that remained a state-of-the-art real-time detector, as well as computing an ‘objectness score’ for bounding boxes using logistic regression, translating as a score of 1 for the anchor box that best overlaps the ground truth, and 0 for other anchor boxes \citep{terven_2023_a}. As well as simplifying the prediction mechanism, this also helps to optimise performance by ensuring only high-objectness-score bounding boxes are considered, reducing the likelihood for false positives. Furthermore, the switch from softmax to the use of binary cross-entropy to train independent logistic classifiers poses the classification problem as a multilabel classification, allowing multiple labels to be assigned to the same bounding box, permitting for overlapping labels (e.g. an object can be a Cat and an Animal). YOLOv3 also allows for multi-scale predictions, making three predictions at three different scales, increasing both its adaptability and performance detecting objects of different sizes \citep{redmon_2018_yolov3}, something extremely applicable to the proposed problem of satellite identification. As such, after gauging the available options, YOLOv3 was selected.

\subsection{YOLO Architecture}
In terms of architecture, YOLOv3 boasts a sizeable feature extractor. Its backbone is referred to as Darknet-53, comprising 53 convolutional layers each with batch normalisation and leaky-ReLU activation. Residual connections link input of the \texttt{1×1} convolutions across the network with the output of the \texttt{3×3} convolutions. The multi-scale detection architecture consists of three elements: \texttt{y1}, \texttt{y2} and \texttt{y3} representing small, medium and large detection scales. YOLOv3 starts with a \texttt{13×13} grid as its initial output at \texttt{y1}, before moving to a \texttt{26×26} grid at \texttt{y2}, merging features from earlier layers by upsampling the \texttt{13×13}  \texttt{y1} grid and concatenating it with the medium-scale features. This process of integration through upsampling then combination is then repeated in \texttt{y3} with a \texttt{52×52}  grid, allowing the algorithm to detect objects of small, medium and large scales in progression. This, therefore, allows YOLOv3 to detect objects of different sizes effectively. For the SPARK dataset used in the current research, with 11 classes, each scale provides an output tensor with a shape of: 

\begin{equation}
N \times N \times \left[ 3 \times (4 + 1 + 11) \right]
\end{equation}
where:
\begin{itemize}
    \item $N \times N$ represents the size of the feature maps $\texttt{y1}$, $\texttt{y2}$, or $\texttt{y3}$.
    \item The factor of $\texttt{3}$ indicates the number of boxes per cell.
    \item The term $(4+1)$ signifies the four bounding box coordinates plus the objectness score.
\end{itemize}

The Darknet-53 backbone and the multi-scale architecture are shown in Figure 2. and Figure 3. respectively.

\begin{figure}
  \centering
  \includegraphics[scale=0.38]{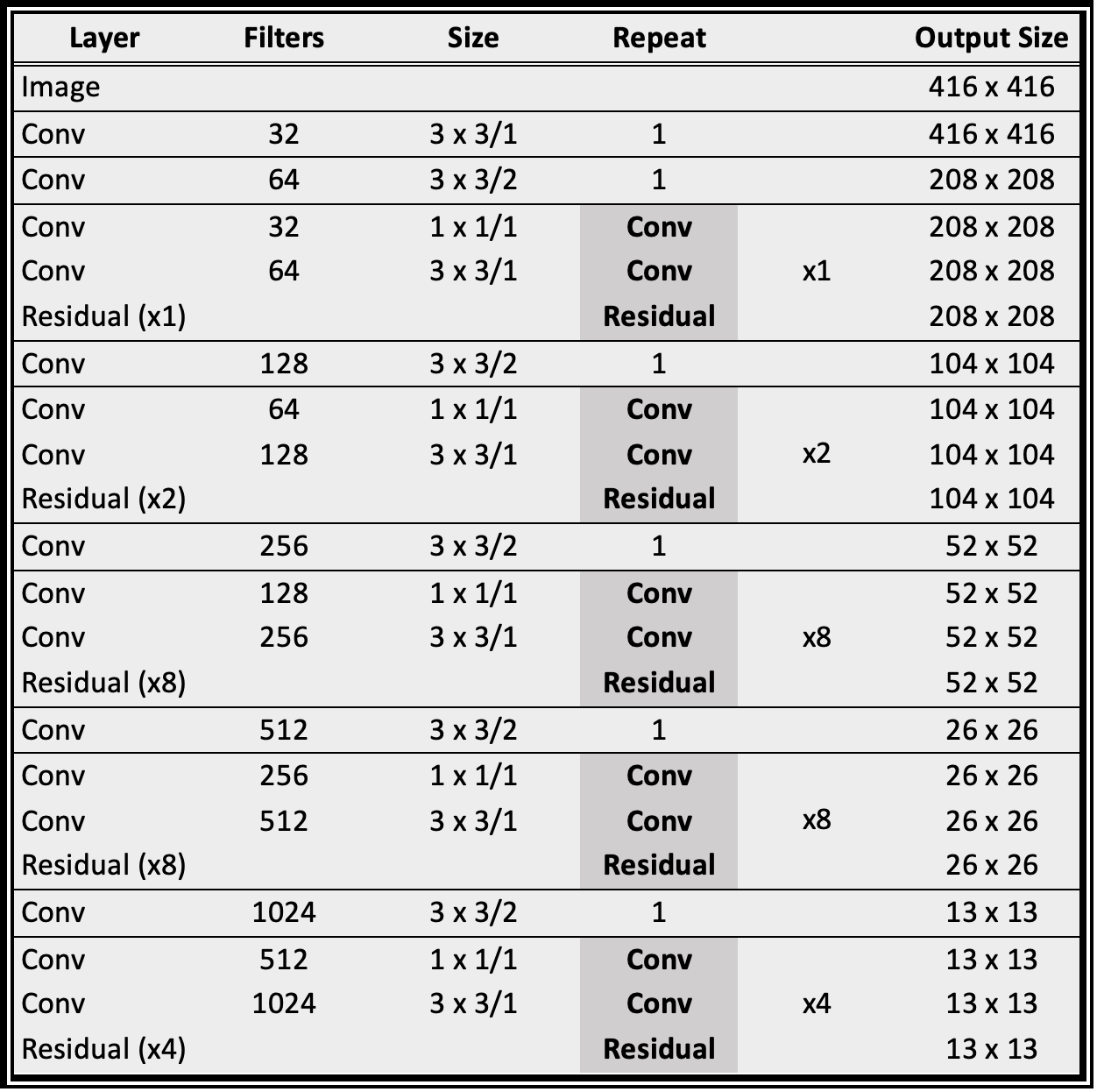}
  \caption{The YOLOv3 Darknet-53 Backbone}
  \label{fig:pendulum}
\end{figure}

\begin{figure*}
  \centering
  \includegraphics[scale=0.4]{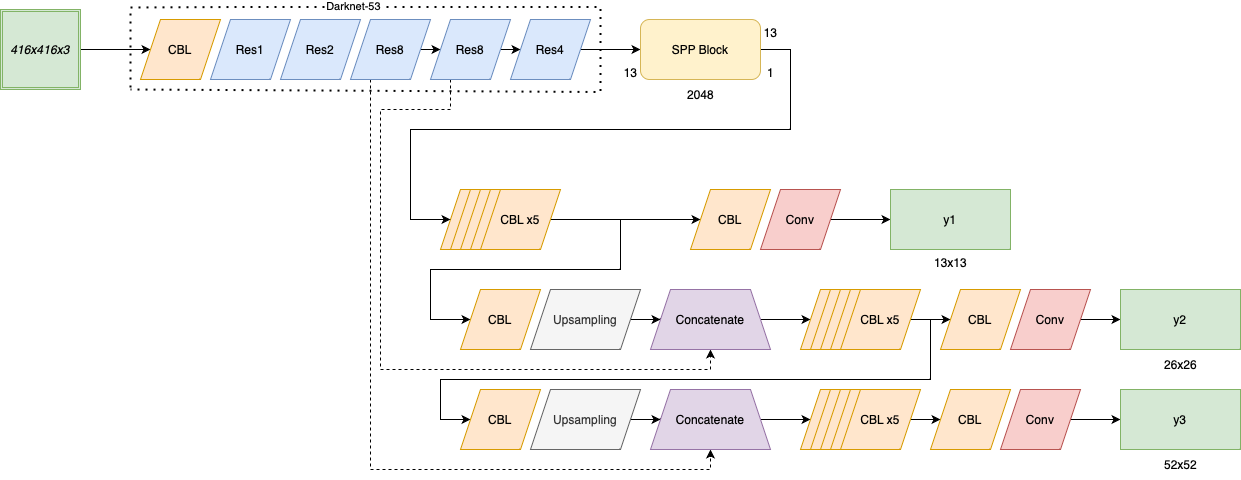}
  \caption{The YOLOv3 Multi-Scale Architecture}
  \label{fig:pendulum}
\end{figure*}

\subsection{UKF}
The responsibility of object tracking was handled by an Unscented Kalman Filter, as discussed above. The UKF, in essence, is a sequence of equations that seek to estimate the most plausible future state of a system \citep{ronghuizhan_2006_neural}. Developed and first published by \citet{julier_2004_unscented}, the UKF built on the limitations of the previous linear KF and Extended Kalman Filter as well as offering a solution to non-linear systems that required KF functionality. The UKF relies on a process known as an Unscented Transformation, a deterministic sampling technique for calculating statistics of a random variable undergoing a non-linear transformation. Instead of linearising the functions, the UT represents them with a set of sample points, known as Sigma points, which can accurately represent both the mean and covariance (uncertainty) values. Two fundamental components of the UKF are the process model, representing the change in system state change over time, and the measurement model, representing how observed measurements relate to system state. These components are crucial to the inner workings of the UKF, which involves two main steps: a prediction step and an update step. In the prediction step, Sigma points are passed through the non-linear process model to estimate state before the predicted mean and covariance are calculated from the propagated points’ weighted average and weighted covariance. In the update step, the predicted points are then passed through the non-linear measurement model, and predicted mean and covariance values are again computed. The cross-covariance between the predicted state and measurement is assessed, determining Kalman gain which proposes the scale of necessary adjustment. The predicted state is then adjusted using the Kalman gain and measurement residual. In design, the scaling parameter and Sigma point weights can be adjusted to influence the spread of Sigma points around the mean, in turn tuning performance.  

\subsection{UKF Architecture}
In terms of application to an image-based problem such as the object tracking of the current study, the UKF works to estimate the state of an object, such as a satellite, as it moves from frame-to-frame. In this sense, the state definition can be represented by the bounding box coordinates for the image, such as those produced by the YOLO output. In the prediction step, Sigma points are generated, representing states the satellite could be in, and propagated through a motion model to predict where the object could be in the following frame. For the current implementation surrounding the SPARK, a constant velocity model was selected. As well as simplifying the estimation process and being the most common design \citep{baisa_2020_derivation}, objects in space that are in stable orbits, such as satellites, are governed by fundamental principles that support the use of a constant velocity model, such as Newton’s Law of Inertia \citep{earman_1973_the} and a related lack of external resistive forces such as atmospheric drag or volatile gravitational interactions \citep{vallado_2014_a}. In the update step, the object is then detected in the next frame, the predicted location is computed from the predicted Sigma points and the UKF adjusts its predicted state by the residual between estimation and measurement. The UKF repeats this process, iteratively, throughout all frames or images, updating its state estimates. Displaying the UKF’s predicted bounding box on the image, next to the actual prediction from an object detection algorithm such as YOLO, can help to visualise the performance of the model on an image-to-image basis. In terms of the current study, the performance of the UKF would be weighed against performance of a linear KF applied to the problem, as employed in the previous research \citep{fitzgerald_2022_space} and offering a good baseline comparison.

\subsection{CosmosDSR}
The current study aims to combine the benefits and functionality of both the YOLOv3 and UKF, proposing a novel algorithm for \textit{Comprehensive Orbital Surveillence and Monitoring Of Space by Detecting Satellite Residuals (CosmosDSR)}. The mathematical representation of the algorithm is presented below.

First, we can assume a typical YOLO output, relating to the predicted bounding box coordinates, and consider this the initial state of the Unscented Kalman Filter at time step \(k\):

\begin{equation}
\hat{x}_k = [x_k, y_k, w_k, h_k, \dot{x}_k, \dot{y}_k, \dot{w}_k, \dot{h}_k]^T
\end{equation}

The relevant velocities \(\dot{x}_k, \dot{y}_k, \dot{w}_k, \dot{h}_k\) are calculated using:

\begin{equation}
\dot{x}_k = \frac{x_k - x_{k-1}}{\Delta t}
\end{equation}

\begin{equation}
\dot{y}_k = \frac{y_k - y_{k-1}}{\Delta t}
\end{equation}

\begin{equation}
\dot{w}_k = \frac{w_k - w_{k-1}}{\Delta t}
\end{equation}

\begin{equation}
\dot{h}_k = \frac{h_k - h_{k-1}}{\Delta t}
\end{equation}

Thus, for predicting the set of sigma points for time step \(k+1\), we can establish the state transition function\footnote{\(b\), a constant coefficient representing the extent that external forces or perturbations \(U_k\) influence the state variables, is set to \(b=0\) for the current application.} \(f\):

\begin{equation}
X_{k+1}^{\prime (i)} = f\left(X_k^{(i)}, U_k\right)
\end{equation}

\begin{equation}
\renewcommand{\arraystretch}{1.5}
f(X_k, U_k) = \begin{bmatrix}
x_k + \dot{x_k} \Delta t + (b^{(x)} \times U_k^{(x)}) \\
y_k + \dot{y_k} \Delta t + (b^{(y)} \times U_k^{(y)}) \\
w_k + \dot{w_k} \Delta t + (b^{(w)} \times U_k^{(w)}) \\
h_k + \dot{h_k} \Delta t + (b^{(h)} \times U_k^{(h)}) \\
\dot{x_k} \\
\dot{y_k} \\
\dot{w_k} \\
\dot{h_k} \\
\end{bmatrix}
\end{equation}

We can denote the first sigma points at time step \(k\), which is equal to the above state estimate at time step \(k\):

\begin{equation}
X_k^{(0)} = \hat{x}_k
\end{equation}

Using the known dimensions of \(n=8\), we can denote the \(i\)th sigma points at time step \(k\) as:

\begin{equation}
X_k^{(i)} = \hat{x}_k + \sqrt{(n+\lambda)} \cdot \mathrm{Chol}(P_k)^{(i)} \quad \forall i = 1,...,n
\end{equation}

\begin{equation}
X_k^{(i)} = \hat{x}_k - \sqrt{(n+\lambda)} \cdot \mathrm{Chol}(P_k)^{(i-n)} \quad \forall i = n+1,...,2n
\end{equation}

Where \(\lambda\) can be defined as:

\begin{equation}
\lambda = \alpha^2 (n + \kappa) - n
\end{equation}
where:
\begin{description}
    \item[\(\lambda\)] is the scaling parameter.
    \item[\(\alpha\)] determines the spread of the sigma points (usually a small positive value).
    \item[\(n\)] is the dimensionality of the state.
    \item[\(\kappa\)] is a secondary scaling parameter (often set to 0 or 3 - n).
    \item[\(Chol\)] is the Cholesky decomposition.
\end{description}

The predicted state at the next time step \(k+1\) can be calculated by taking the weighted average of the propagated sigma points:

\begin{equation}
\hat{x}_{k+1} = \sum_{i=0}^{2n} w_m^{(i)} X_{k+1}'^{(i)}
\end{equation}

where:
\begin{description}
    \item[\(W_m\)] represents the mean weights. 
\end{description}

The predicted covariance at this time step can be defined:

\begin{equation}
P_{k+1|k} = \sum_{i=0}^{2n} w_c^{(i)} \left( X_{k+1}'^{(i)} - \hat{x}_{k+1} \right) \left( X_{k+1}'^{(i)} - \hat{x}_{k+1} \right)^T + Q_1
\end{equation}

where:
\begin{description}
    \item[\(W_c\)] represents the covariance weights.
    \item[\(Q_1\)] represents the process noise covariance matrix. 
\end{description}

The mean weights \(Wm\) and covariance weights \(Wc\) can be defined as:

\begin{equation}
w_m^{(0)} = \frac{\lambda}{n + \lambda}
\end{equation}

\begin{equation}
w_c^{(0)} = \frac{\lambda}{n + \lambda} + (1 - \alpha^2 + \beta)
\end{equation}

\begin{equation}
w_m^{(i)} = w_c^{(i)} = \frac{1}{2(n + \lambda)}
\end{equation}

where:
\begin{description}
    \item[\(\beta\)] provides prior knowledge about the state distribution.
\end{description}

We can then transform the predicted sigma points into the measurement space:

\begin{equation}
Z_{k+1}^{(i)} = h\left(X_{k+1}^{\prime (i)}\right) = H \cdot X_{k+1}^{\prime (i)}
\end{equation}

Where the measurement function \(H\) can be defined as:

\begin{equation}
H = 
\begin{bmatrix}
1 & 0 & 0 & 0 & 0 & 0 & 0 & 0 \\
0 & 1 & 0 & 0 & 0 & 0 & 0 & 0 \\
0 & 0 & 1 & 0 & 0 & 0 & 0 & 0 \\
0 & 0 & 0 & 1 & 0 & 0 & 0 & 0 \\
\end{bmatrix}
\end{equation}

Now, the predicted measurement mean can be calculated using the transformed sigma points:

\begin{equation}
\hat{z}_{k+1} = \sum_{i=0}^{2n} w_m^{(i)} Z_{k+1}^{(i)}
\end{equation}

As can the measurement covariance:

\begin{equation}
P^z_{{k+1|k+1}} = \sum_{i=0}^{2n} w_c^{(i)} \left( Z_{k+1}^{(i)} - \hat{z}_{k+1} \right) \left( Z_{k+1}^{(i)} - \hat{z}_{k+1} \right)^T + Q_2
\end{equation}

where:
\begin{description}
    \item[\(Q_2\)] represents the measurement noise covariance matrix. 
\end{description}

Following this, the cross-covariance between the state and measurement can be calculated:

\begin{equation}
P^{xz}_{{k+1|k+1}} = \sum_{i=0}^{2n} w_c^{(i)} \left( X_{k+1}'^{(i)} - \hat{x}_{k+1} \right) \left( Z_{k+1}^{(i)} - \hat{z}_{k+1} \right)^T
\end{equation}

The Kalman gain \(K\) can then be computed:

\begin{equation}
K_{k+1} = P^{xz}_{k+1|k+1} \left( P^{z}_{k+1|k+1} \right)^{-1}
\end{equation}

Finally, we can update the state estimate and the covariance using the new measurements:

\begin{equation}
X_{k+1} = \hat{x}_{k+1} + K_{k+1} (y_{k+1} - \hat{z}_{k+1})
\end{equation}

\begin{equation}
P_{k+1|k+1} = P_{k+1|k} - K_{k+1} P^{z}_{k+1|k+1} K_{k+1}^T
\end{equation}

\subsection{Implementation}
One of the major benefits in using YOLOv3 is the amount of research, implementations and support available to augment development. Specifically, the current research elected to use the Darknet YOLOv3 wrapper developed by Alexey Bochkovskiy and available in his GitHub repository \citep{alexey_2020_alexeyabdarknet}. This is an extremely popular fork of the original Darknet repository, offering a wrapper that enhances \citet{redmon_2018_yolov3}'s YOLOv3, offering improvements in speed, stability and flexibility as well as integrated customisability. Although primarily designed to be used in C, Python functionality is included and, as such, the wrapper can be used within Google Colab.  In order to tune relevant hyperparameters, a GridSearch was used. This involved systematically exploring a range of values for a set of defined hyperparameters as to identify which combination of values result in the best performance on a small test set of relevant images, thus the optimal configuration \citep{liashchynskyi_2019_grid}. This was applied to a subset of the SPARK data consisting of n=1000 images which was split into train \texttt{(n=800)} and validate \texttt{(n=200)} sets. This data was used to run YOLOv3 for each set of parameters, \texttt{n=200} iterations per combination, before calculating average loss to approximate performance. The best parameters were then extracted, giving optimal values of \texttt{[learning rate: 0.001, momentum: 0.9, decay: 0.0005]}. The YOLO output information, including class predictions and bounding box information, of the eventual trained YOLO algorithm when presented with an image would be extracted and saved to a file to be used by the UKF for object tracking.\newline

The UKF for the current study was developed using the \textit{filterpy} library, a comprehensive toolkit designed for filter and state estimation processes within Python, published by Roger Labbe and available in his GitHub repository \citep{labbe_2023_filterpy, labbe_2023_rlabbefilterpy}. This specific implementation was chosen as it offers a robust and optimised version of the core structure of both necessary KF methods required in the current research. Regarding the current study’s KF code, developed using Python within Google Colab, the use of the \textit{filterpy} library allowed for efficient customisation and flexibility to tailor the filter parameters to the current application. The UKF was initialised with Merwe-Scaled Sigma Points, to preserve the Gaussian characteristics of the system’s state whilst offering high stability, accuracy and performance \citep{vandermerwe_sigmapoint}; the measurement and process noise covariance matrices were initialised as identity matrices scaled to an arbitrary value of \texttt{0.1} for neutrality and simplicity \citep{formentin_2014_an}. The process for designing the linear KF was mostly identical. The state transition function \(f\left(X_k^{(i)}, U_k\right)\) and measurement function \(H\)  were designed in line with the selected constant velocity model \citep{popoli_1999_design}, and the time delta \(dt\) was adjusted to an optimal value of \texttt{0.01}. The design of the code interfaced with the previous YOLO algorithm, post-training, to detect the bounding box for an image before passing that image through both a linear KF and UKF algorithm to predict the state of the satellite in the following image. This was then repeated with each image in sequence, before mean squared error (MSE) and root mean squared error (RMSE) between the predicted state and the detected bounding box were calculated for each algorithm, thus allowing for the performance over that set of images to be compared and contrasted. The creation and displaying of images was handled with the cv2 and matplotlib packages \citep{bradski_2020_the, hunter_2007_matplotlib} and the generated images, containing both bounding boxes, were also saved to a data-specific folder for visual inspection.

\section{Method}
\subsection{YOLO Pre-Training}
The first task of the current study was to train the YOLOv3 algorithm to correctly detect and classify the eleven categories of satellite experienced within the SPARK dataset. Stream-1 images were split into training and validation sets at a ratio of 4:1, resulting in \texttt{n=17600} images with associated annotations in the train set and \texttt{n=4400} images with annotations in the validation set. After building Darknet within Colab, the train command was invoked, setting the parameters to match the prepared .data and .cfg files, setting the initial weights to the downloaded darknet53.conv.74 weights file. The YOLOv3 algorithm was trained for one epoch of \texttt{n=22000} iterations within Colab using an NVIDIA Tesla V100-SXM2-16GB GPU. Following successful completion of training, the map command was invoked using the generated weights file. Finally, a set of varied test images of the 11 classes, taken from other SPARK Stream-1 data not used in the training, was uploaded and the test command was invoked on each to visually gauge performance of the YOLO.

\subsection{CosmosDSR Implementation}
With the successful training of the YOLOv3 algorithm, the generated weights file would be used for all Darknet initialisations going forward. The next aim of the research was to implement a UKF to track the position of RSOs between consecutive frames of the Stream-2 data. The UKF code was executed on one of the previously elected Stream-2 sets of consecutive images. For each image, Darknet was used to generate a bounding box prediction using the trained weights file, before storing this information to a file. The UKF code then read and parsed this information to give definitive bounding box coordinates, which would be displayed in green on the image. The state estimation from the UKF was displayed as a red box on the image, and this was saved to a nominal folder in sequence. Following the application of the UKF to all images in the set, this was repeated with the linear KF, generating another folder of output images. After both algorithms had been applied, the MSE and RMSE for both was calculated. The MSE can be thought to represent the squared difference between the KF estimated and YOLO predicted values, an important metric for algorithm cross-evaluation due to its inherent method of penalising larger errors more than smaller errors \citep{hodson_2021_mean}. However, this is hard to intuitively interpret and, as such, computing the RMSE, representing the average prediction error in terms of pixels \citep{ajala_2022_comparing}, allows the results to be both understandable and relatable. This process was then repeated with further Stream-2 image sets. 

\section{Results}
\subsection{YOLOv3 Classification}
Following training, the YOLOv3 algorithm returned a Mean Average Precision (mAP) score, a current benchmark metric used within CV research to establish the performance of a model with regards to making accurate predictions \citep{padilla_2021_a}, of \texttt{97.18\%} at a standard Intersection over Union (IoU) threshold of \texttt{0.5} across the validation set. The average IoU value was \texttt{83.03\%}, indicating accurate object localisation. Furthermore, an F1-score, the harmonic mean of precision and recall scores \citep{hicks_2022_on}, of \texttt{0.95} was achieved. Overall, there were \texttt{n=4163} true positives, \texttt{n=209} false positives and \texttt{n=237} false negatives across all classes. Class-wise prediction statistics are shown below in Table 2.\newline

\begin{table*}
\centering
\caption{YOLOv3 class-wise prediction statistics after one epoch}
\begin{tabular}{|c|l|c|c|c|}
\hline
\textbf{Class ID} & \textbf{Class Name} & \textbf{Average Precision (AP\%)} & \textbf{True Positives} & \textbf{False Positives} \\
\hline
0 & lisa\_pathfinder & 99.94 & 361 & 2 \\
1 & proba\_data3\_csc & 98.30 & 370 & 3 \\
2 & smart\_1 & 96.66 & 365 & 22 \\
3 & xmm\_newton & 94.55 & 375 & 58 \\
4 & soho & 96.49 & 400 & 8 \\
5 & earth\_observation\_sat\_1 & 90.02 & 369 & 28 \\
6 & debris & 99.26 & 421 & 4 \\
7 & proba\_2 & 98.81 & 364 & 9 \\
8 & proba\_3\_ocs & 98.06 & 403 & 66 \\
9 & cheops & 98.56 & 374 & 5 \\
10 & double\_star & 98.31 & 361 & 4 \\
\hline
\end{tabular}
\end{table*}

The observed results indicate a high accuracy across all classes, suggesting excellent performance of the YOLOv3 algorithm on the contemporary SPARK data. The models adeptness at classifying RSO’s, as well as localising them correctly within an image, suggests robust detection capabilities and signifies that the first aim of the current research was a success. An example of visual output, where YOLO correctly predicts distinct RSO classes with high certainty, is shown in Figure 4. 

\begin{figure}
  \centering
  \includegraphics[scale=0.6]{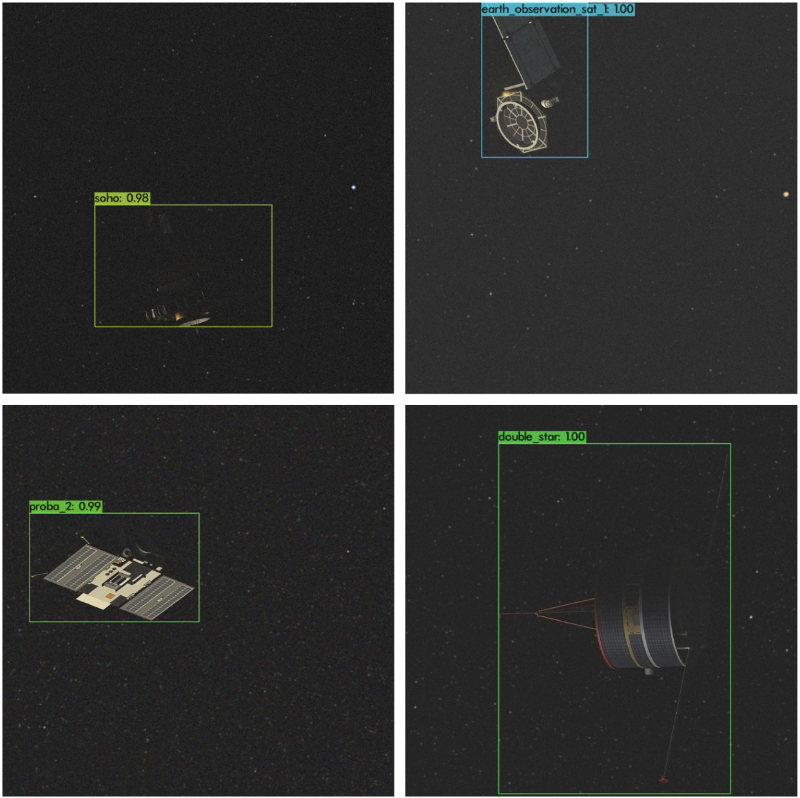}
  \caption{YOLOv3 test output of Stream-1 images}
  \label{fig:pendulum}
\end{figure}

\subsection{CosmosDSR Performance}
On the initial Stream-2 validation test, the YOLO performed well, once again predicting the correct class with 99\% certainty, as can be seen in Figure 5.

\begin{figure}
  \centering
  \includegraphics[scale=0.6]{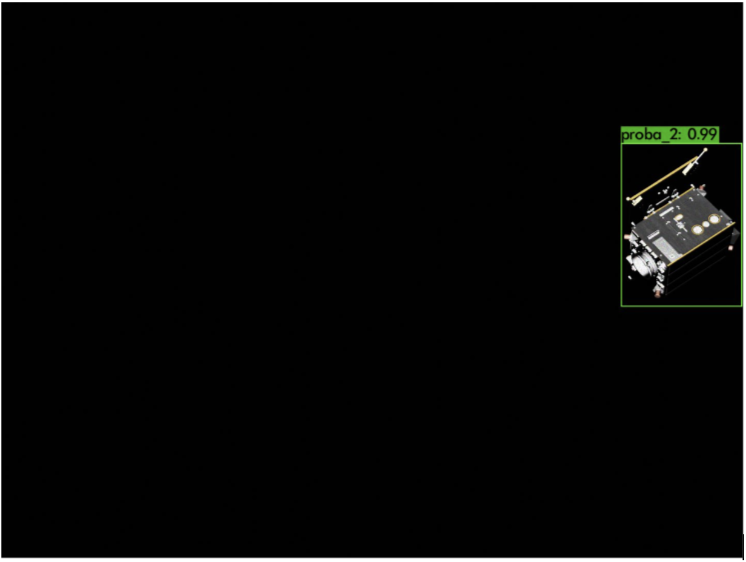}
  \caption{YOLOv3 test output of a Stream-2 image}
  \label{fig:pendulum}
\end{figure}

This suggests the algorithm was able to generalise relatively well and further stands to uphold the success of its implementation. The UKF code was then executed on each of the Stream-2 sets of images, results of which are shown below in Table 3. 

\begin{table*}
\centering
\caption{Evaluation of the UKF/KF on Stream-2 image sets (* at a value lower than 2dp)}
\begin{tabular}{|l|c|c|c|c|l|}
\hline
\textbf{Data} & \textbf{UKF MSE} & \textbf{UKF RMSE} & \textbf{LKF MSE} & \textbf{LKF RMSE} & \textbf{Best Performance} \\
\hline
GT086 & 1.88 & 1.37 & 1.88 & 1.37 & UKF* \\
GT066 & 1.86 & 1.37 & 1.86 & 1.37 & UKF* \\
GT011 & 4.15 & 2.04 & 4.17 & 2.04 & UKF \\
GT053 & 3.55 & 1.88 & 3.55 & 1.88 & LKF* \\
GT047 & 2.73 & 1.65 & 2.74 & 1.65 & UKF \\
Mean $(\mu)$ & 2.83 & 1.66 & 2.84 & 1.66 & UKF \\
\hline
\end{tabular}
\end{table*}

The results imply that the UKF performed well when applied to track the SPARK RSOs across consecutive Stream-2 images. Overall, the results indicate that the UKF did outperform the linear KF for four image sets, however the linear KF also performed well, and the disparity was minimal. Computed RMSE results for all experiments, as shown in Table 3., were low \((\mu = 1.66)\) with differences generally being minute. Considering the \texttt{440x1080} resolution of the Stream-2 images, the results were sufficient to conclude that in the current research, the UKF algorithm was able to accurately track a satellite; and when combined with the SPARK-trained YOLOv3 into a novel algorithm, function as a relatively strong object detection and tracking model. Although the sample size of image sets was small, the results across each of the tests are sufficiently congruent to theorise that further application to similar data would result in similar results. A panel, showing the first sixteen generated images of the UKF execution on the GT086 dataset is shown below in Figure 6. This is followed by a similar panel of images from the linear KF execution on the same data in Figure 7. The predicted bounding box from the YOLO is shown in green and the object state estimation from the UKF/LKF is shown in red. 

\begin{figure*}
  \centering
  \includegraphics[scale=0.8]{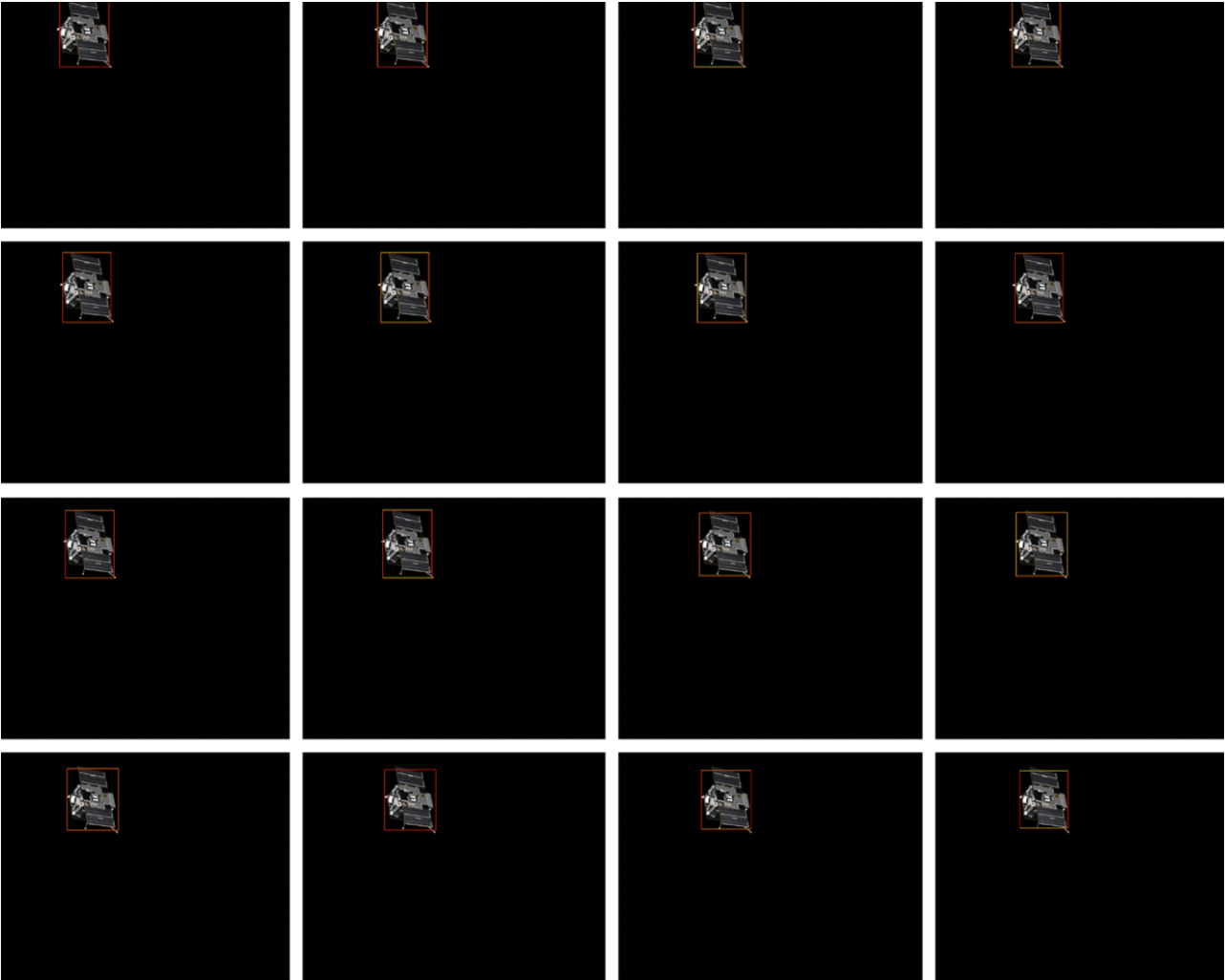}
  \caption{The first 16 generated images from the UKF application on the GT086 image set. Green bounding box represents YOLO prediction, red bounding box represents UKF estimation.}
  \label{fig:pendulum}
\end{figure*}

\begin{figure*}
  \centering
  \includegraphics[scale=0.8]{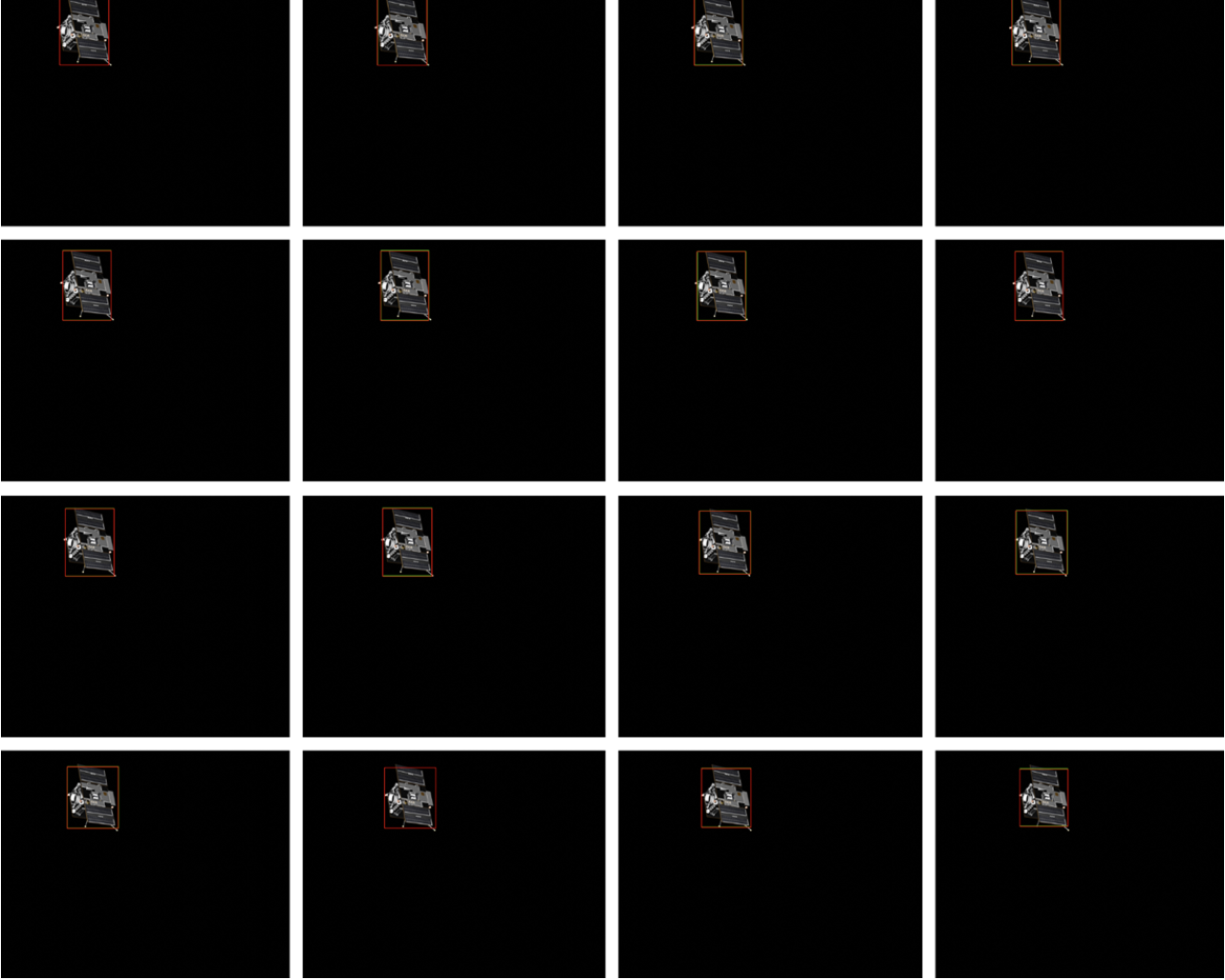}
  \caption{The first 16 generated images from the LKF application on the GT086 image set. Green bounding box represents YOLO prediction, red bounding box represents UKF estimation.}
  \label{fig:pendulum}
\end{figure*}

Upon reviewing the output images from both the UKF and LKF, apart from some small disparity, it is hard to visually distinguish between the two. Both algorithms exhibited good performance in tracking the satellites and this mirrors the achieved MSE/RMSE results. Visually, the KF state estimate bounding boxes mostly overlap the predicted OD bounding boxes, suggesting the filters’ predictions strongly align with the YOLO detections and signifying both the consistency and reliability of the combined approaches. Overall, despite such similarities, and being cognisant of the limited number of experiments completed, the UKF did outperform the linear KF with regards to SPARK RSO tracking in consecutive images.

\section{Discussion}
The end performance of the algorithm is indicative that hyperparameter values were appropriately selected. It may be of some use for future studies to include extra variables into the tests such as dropout, batch size, subdivisions or the number of epochs. As the code is already prepared, including these would need minimal further work and should be considered in any resulting future research, as it would also further increase the efficiency of the final algorithm.  

The performance of the YOLOv3 itself was also exemplary, with results suggesting the algorithm could accurately classify all eleven categories of satellite/RSO correctly with minimal issues such as false positives or negatives. A paper by \citet{razavian_2014_cnn} supports the idea that pre-trained networks can learn universally useful features, highlighting powerful ‘off-the-shelf’ features from networks trained on ImageNet. In the current study, the pre-trained weights file seemed to be a good starting point for the SPARK-focused training, with the favourable results suggesting that the model converged quickly; this in turn may suggest that transfer learning may have occurred. Furthermore, it can be inferred that the SPARK data itself helped to facilitate the training performance as it is a well-established and well-developed dataset that has already been successfully applied to ML tasks \citep{musallam_spark}. The clarity of the data may have led to the certainty of the YOLO when predicting images with values of \texttt{>0.99\%}, which is rarely observed. This can possibly be attributed to the clear categorisation of the data, as well as the fact it has been synthetically generated and as such, has minimal levels of noise and occlusion. Such drawbacks of YOLO, as outlined in the previous literature \citep{terven_2023_a, barreiros_2021_zebrafish}, were therefore not sufficiently tested. This is something that must be explored in future research; according to Zheng et al. \citep{zheng_2015_an}, astronomical images have higher proportions of noise, large dynamic ranges of intensities and often unclear boundaries when compared to ordinary ‘non-Space’ images. Undoubtedly, the organic conditions of Space are also complex and in many cases hostile, making clear imaging of satellites in any way that closely resembles the SPARK impractical, which may somewhat explain the lack of easily retrievable real images observed when selecting the data for the current project. Therefore, although the YOLO’s performance on the test set of the SPARK images was good, applying this to real scenarios may not be straightforward. \citet{seemakurthy_2022_domain} explain that although algorithms may perform well on similar data, problems with generalisation, the application of a specific model to a wider problem domain, remain a concern. With regards to RSO classification, an exceedingly complex and ever-changing problem that contains far more classes and designations of objects than can be expressed by even a comprehensive synthetic dataset such as the SPARK, a lack of generalisation quickly negates the validity of the approach. However, the ability of the algorithm to accurately predict the satellite in the Stream-2 images, despite the resolution of the image increasing and the object size being mostly vastly different from the Stream-1 images used for training, suggests the YOLO was at least able to generalise to different conditions even if the subject remained the same. Future research could apply the weights generated by the current study to real images of space, or another synthetic dataset, as to assess the model’s ability to generalise to the broader domain of RSO classification. Comparing the performance to that of the method outlined by \citet{fitzgerald_2022_space}, the current YOLOv3 resulted in an F1-score of \texttt{0.95}, which is larger than the F1-score of \texttt{0.81} observed by the author’s detection algorithm on their main dataset. Additionally, the current study’s algorithm performed as good or better when compared to all tested classical algorithms –  SIFT, HOGOR and FF – across all data subsets, suggesting at least comparable if not better performance of the current YOLOv3 model. Again, it is important to note that the current research focused on a non-identical, simpler problem with fewer experiments and as such, any comparison can only be thought of in rough terms despite regarding a similar domain. Nevertheless, the obtained results suggest the current YOLOv3 architecture is an appropriate algorithm for RSO identification and classification, offering a new angle to approach the problem and establishing a base for further development.

The second stage of the project, integrating the trained YOLO into a novel solution for RSO tracking through the use of UKF, was also completed with relative success. Building from the future research by \citet{fitzgerald_2022_space} and \citet{barreiros_2021_zebrafish}, in which linear Kalman filters were combined with YOLO algorithms for such purposes, the current research elected to use UKF. This mediated the evaluation of another algorithm, whilst allowing for both internal and external comparisons through equating performance to both the previous literature and a concurrently developed LKF in the current study. Following the design of the code, the performance of both the UKF and LKF on the five sets of Stream-2 data was gathered, with results indicating good performance of both algorithms on each dataset. Across the five sets, the average RMSE for the UKF was \textbf{}1.66 and the average RMSE for the LKF was also \textbf{}1.66. The differences between the results for both algorithms, both numerically and visually, were extremely small and indicated respectable object tracking of the RSO across the consecutive images. Furthermore, when the model was unable to make a prediction with YOLO due to uncertainty or loss of object, the KF still accurately estimated the position across images, correctly indicating that a satellite is in the area. The accuracy of such predictions is reflected in the RMSE – by trait a measure susceptible to skewed values due to small amounts of high-value outliers \citep{chai_2014_root} – thus even a single bad prediction could drastically increase the error. In this vein, RMSE can be thought of as a robust measure of object tracker performance comparisons that is used in many studies within the literature \citep{zhao_performance, wahn_2023_predicting, sunindyo_2020_traffic}, thus a low value for both UKF and LKF signifies not only good but, comparatively, analogous performance. With regards to the main, internal comparison between the two algorithms, for the current application, either would be a suitable choice. However, the UKF overall did slightly outperform the LKF on the test sets, and although such disparity is minute, in a complex domain such as orbital surveillance even small increases in performance or efficiency may be important. This comparison should be expanded on by future research into the area as to test the algorithms’ performances on larger samples of similar yet more problematic tasks, to ascertain how the disparity changes when some non-linearity is introduced. The current study focused on a relatively simple, linear problem and as such, and as shown by the results, the LKF would have been sufficient. However, as explained by \citet{kaineg_2020_the} and substantiated by \citet{chen_2011_the}, the problem of orbital debris is multifaceted, everchanging and extremely challenging, thus encountering non-linearity is almost certain whilst attempting to mitigate it. In such cases, the advantages of using the UKF would be much more apparent, and good performance on the linear problem within the current research implies that the algorithm would be well suited to the task. The results stand to uphold the sentiments made by the authors of the previous literature; that adopting deep-learned detection and tracking algorithms are extremely beneficial for improving our understanding and proficiency with object identification. The YOLOv5/LKF combination applied by \citet{fitzgerald_2022_space} generated results that significantly improved on previous attempts, with the author describing the research as a success and \citet{barreiros_2021_zebrafish}’s application of a YOLOv2/LKF combination proposed a similarly effective method for a distinctly unique problem. As the YOLOv3/UKF architecture utilised in the current study was designed to build upon such research, testing a new combination of algorithms, the favourable results suggest it too was a success. Obviously, as with the previous literature, the research was not without its limitations.

One such limitation may arise from the lack of ability to validly evaluate. Whilst a comparison can be made to the previous studies, in the sense that all surrounded a novel application of an object detection and tracking algorithm to a problem with a high rate of success, making a direct quantitative assessment is not possible as all propose a different evaluation metric. As previously stated, the current study used MSE/RMSE to evaluate performance due to benefits of robustness, outlier detection and simplicity of internal comparisons. \citet{fitzgerald_2022_space}, however, evaluated tracking performance using solely F1-score. This has the benefit of balancing recall and precision scores, which is crucial for balancing tracked objects and minimising missed predictions within object tracking tasks \citep{song_2022_performance}. However, F1-score treats each image or frame independently and does not give a good representation of the temporal continuity of the tracking, thus it does not measure the fluidity of the algorithm’s application across data and may disregard crucial information about its performance. As such, researchers often provide multiple metrics to bypass such issues, such as offering F1-score alongside RMSE \citep{chen_2021_dualmodality}. Future research should endeavour to calculate and present sufficient metrics to evaluate performance from all angles. \citep{barreiros_2021_zebrafish} did not provide either score; due to the uniqueness of the problem, the authors offered bespoke metrics regarding the number of frames tracked correctly for a single target and the probability for correct target re-identification following occlusion. As such, no quantitative comparison is possible. Another limitation may arise from the Stream-2 data used for the object tracking. Whilst the SPARK dataset as a whole has proved a good asset for the research, helping to correctly train the YOLO on eleven distinct classes contained within the Stream-1 data, the sets of images used for the KF all contained the same satellite, proba2, indicating class imbalance. Problems with imbalance within object detection applications are already well documented \citep{oksuz_2020_imbalance, johnson_2019_survey}, such as reduced versatility and limited generalisation which negatively impacts the applicability of the model. Future research must build on this by applying the novel model not only to real data as to ensure it translates correctly \citep{martingeorgljungqvist_2023_object}, but also to Stream-2 sets of other classes as to assess KF performance on different satellites. However, as detection accuracies were high for all classes, there is no reason to assume that the KF algorithms could not accurately track any of them with similar success to its performance detecting \textit{proba2}.

\section{Conclusion}
In conclusion, the current study aimed to design a novel object detection and tracking model, comprising a YOLOv3 object detector trained on SPARK RSO image data combined with a UKF algorithm to facilitate object tracking across consecutive images, as to offer a solution to automatically track orbital debris. Results indicated that the YOLOv3 was able to accurately detect and classify satellites into eleven distinct categories found within the SPARK, with minimal errors such as false positives or negatives. Furthermore, the UKF was able to successfully track a satellite throughout sets of consecutive images with good performance. This was directly compared to an LKF, which also performed well, with results being extremely close and UKF displaying only a slight improvement. Interpreting the results suggests the study, based on its aims, was a success and the novel YOLOv3/UKF algorithm developed can be considered an appropriate solution for RSO detection and tracking. However, future research should attempt to gather non-synthetic images and apply the model to a larger sample of tasks, as well as diversifying the subject of images used within the KF predictions. Additionally, all future research should aim to give a more comprehensive array of performance evaluation metrics as to be able to more easily externally validify the results with previous literature. Thought must be given to incorporating the model with debris mitigation strategies. Finally, the issues of occlusion and noise need to be investigated as to further assess their significance on performance in a domain where they are regularly encountered. Accomplishing the above would supplement the findings of the current study, increase the knowledge of the research area and ultimately move towards finding a solution to the Kessler syndrome, cleaning up the stellar environment and facilitating future space exploration. 

\section*{Acknowledgments}
Thanks to Dr. Djamila Aouada and the SPARK Organisation Team at the University of Luxembourg, specifically Dr. Arunkumar Rathinam for facilitating access to the SPARK dataset and mediating further queries relating to its usage.

Further thanks to Dr. Vijayan Asari of the University of Ohio for providing information surrounding the previous research.

The data that support the findings of this study are available from the corresponding author upon reasonable request.

\bibliography{elsarticle-jasr-template/refs.bib}
\bibliographystyle{jasr-model5-names}
\biboptions{authoryear}
\end{document}